\documentclass[twocolumn,secnumroman,amssymb, amsmath, nofootinbib,
nobibnotes, aps, prb, showpacs]{revtex4}

\usepackage{graphicx}
\usepackage{dcolumn}
\usepackage{amsmath}
\usepackage{here}%

\begin{document}

\preprint{HEP/123-QED}

\title{Disappearance and Survival of Superconductivity in FeSe under High Pressure
}
\author{Kiyotaka \textsc{Miyoshi}$^{1,2}$, Shota \textsc{Yamamoto}$^1$, Atsushi \textsc{Shiota}$^1$, 
Takuya \textsc{Matsuoka}$^1$, Masaki \textsc{Ohe}$^1$, \\ 
Yumi \textsc{Yamamoto}$^1$ and Shijo \textsc{Nishigori}$^3$} 
\affiliation{%
$^1$Department of Physics and Materials Science, Shimane University, Matsue 690-8504, Japan \\
}%
\affiliation{%
$^2$Next Generation TATARA Co-Creation Centre, Shimane University, Matsue 690-8504, Japan \\
}
\affiliation{%
$^3$Department of Materials Analysis, CIRS, Shimane University, Matsue 690-8504, Japan 
}

\date{\today}

\begin{abstract}
Superconductivity in FeSe was investigated under high pressure through the measurements 
of DC magnetization by using a diamond anvil cell. 
We successfully observed that the disappearance of the superconductivity as a results of 
the appearance of the non-superconducting ortho I\hspace{-1pt}I phase 
above $\sim$7 GPa ($\sim$5 GPa), when Ar (glycerin) is used as the pressure transmitting media. 
Contrary to this, it has been found that the superconductivity even survives under pressure above 7 GPa, 
when the thickness ($t$) of a platelet-single crystal specimen is reduced. 
The survival of the superconductivity above 7 GPa is consistent with a previous observation 
under hydrostatic pressure by using a cubic anvil apparatus, suggesting that the hydrostaticity 
of the pressure is improved by reducing $t$. It is also inferred that 
the appearance of the ortho I\hspace{-1pt}I phase is due to uniaxial stress along the [001] direction.
\end{abstract}


\maketitle
Since the discovery of superconductivity in LaFeAsO$_{\rm 1-x}$F$_{\rm x}$\cite{kamihara}, 
intensive studies over the past decade have uncovered a wide variety of 
iron-pnictide superconductors and common unconventional superconductivity 
emerged from the competition with the antiferromagnetic (AFM) phase 
and electronic nematic phase with an orthorhombic structure, 
which is inferred from the characteristic phase diagrams. 
In early studies, a typical phase diagram for AFe$_2$As$_2$ (A=Sr, Ba, Ca, Eu) has been established, where 
a superconducting dome appears with disappearance of both AFM and 
nematic phases by tuning carrier doping\cite{rotter,ni,sefat,kasahara2} 
or applying external 
pressure\cite{tori,ishikawa,colo,yamazaki,kotegawa1,matsubayashi,terashima1}. 
In addition to a similar competition between the superconductivity and 
AFM (orthorhombic) phase in LaFeAsO$_{\rm 1-x}$F$_{\rm x}$\cite{luetkens},  
the coexistence of superconductivity with these phases has been found for 
NaFe$_{\rm 1-x}$A$_{\rm x}$As (A=Co\cite{parker,wang1}, Cu\cite{wang2}) 
and SrVO$_3$FeAs\cite{ueshima,holenstein}. 
In addition, a twin-dome structure and plateau-like single-dome structure with bipartite parent phases 
were found in RFeAsO$_{\rm 1-x}$H$_{\rm x}$ with R=La\cite{iimura} 
and Sm\cite{hiraishi}, respectively. In comparison, 
LaFeAs$_{\rm 1-x}$P$_{\rm x}$O was found to exhibit two superconducting domes 
separated by an AFM phase\cite{cwang,lai,mukuda},  
suggesting that a different superconducting state is realized in each dome. 
To establish the phase diagram for various iron-pnictide 
superconductors is crucial to throw further light on the superconducting mechanism, because 
certain fluctuations arising from the ordered states that are either competing or 
coexisting with the superconductivity in the phase diagram are promising candidates 
for paring glue. 

One of the most attractive subjects to explore the temperature ($T$)-pressure ($P$) 
phase diagram is FeSe. This is because 
the superconducting transition temperature $T_{\rm c}$ of FeSe 
was found to be highly enhanced in early studies\cite{masaki,medvedev,margadonna,braithwaite,
miyoshi09}. Furthermore, the nematic state was observed to 
appear without accompanying AFM order at ambient pressure, suggesting that FeSe provides 
an interesting test ground for examining the competition and coexistence of 
the superconducting, AFM, and nematic phases. Indeed, the disappearance of the 
nematic transition and a three-step increase in $T_{\rm c}$ under pressure 
was reported.\cite{miyoshi14} However, more intensive studies under pressure 
succeeded in elucidating the $T$$-$$P$ phase diagram
since single crystal growth to obtain purely tetragonal FeSe without 
mixing of hexagonal phase was enabled\cite{bohmer2,bohmer3}. 
A pressure induced AFM phase above 1.2 GPa has been shown to exist\cite{terashima2} 
and a $T$$-$$P$ phase diagram where a dome-shaped AFM phase competes with 
two other phases has been reported.\cite{sun} 
The AFM ordering was shown to be of the stripe-type 
through NMR\cite{wang3} and $\mu$SR\cite{khasanov1} measurements and 
also to merge with the structural (nematic) transition above 1.7 GPa.\cite{kothapalli} 
This suggests that the AFM and nematicity are strongly coupled with each other in FeSe under pressure, 
similar to other iron-based superconductors.

An important feature of the $T$$-$$P$ phase diagram, 
which was determined by the measurements of electrical resistivity and 
AC susceptibility using a cubic anvil apparatus (CAA), is 
that $T_{\rm c}$ is significantly enhanced and reaches a maximum of 
$\sim$37 K at $\sim$6 GPa, above which the AFM phase boundary line is 
suddenly terminated, demonstrating the competing nature of the AFM and 
superconducting phases.\cite{sun} In addition, recent X-ray diffraction (XRD) 
measurements detected a mixing of an orthorhombic $Pnma$ phase with 
MnP-type structure (ortho I\hspace{-1pt}I) above 6 GPa, 
which is non-superconducting and characterized by a three-dimensional 
network of face sharing FeSe$_6$ octahedra, 
with a superconducting orthorhombic $Cmma$ phase (ortho I).\cite{svitlyk} 
A similar transition to the ortho I\hspace{-1pt}I phase was also observed 
through measurements of XRD\cite{bohmer4} and X-ray absorption spectroscopy\cite{lebert}. 
FeSe was known to undergo a structural transition at high pressure 
before the synthesis of purely tetragonal FeSe is possible.\cite{medvedev,margadonna,braithwaite,kumar} 
Although the superconductivity would be expected to 
disappear as the result of the appearance of the ortho I\hspace{-1pt}I phase, 
no signature of the disappearance was inconsistently observed in AC susceptibility 
measurements using a CAA up to $\sim$9 GPa.\cite{sun} 
It is important to clarify the origin of the inconsistency 
to gain more insight into 
the superconductivity of FeSe and also for further investigations under pressure. 
The origin is likely to be attributable to the difference in the 
hydrostaticity of the pressure cell, considering that diamond anvil cells (DACs) were 
commonly used for the observation of 
the ortho I\hspace{-1pt}I phase.\cite{svitlyk,bohmer4,lebert}

In the present work, we performed DC magnetization measurements under pressure 
for single crystals of FeSe using DAC and Ar as the pressure-transmitting medium (PTM)
and investigated the pressure variation of 
the superconducting volume fraction. We successfully observed that 
the superconductivity disappears and survives under pressure above 7 GPa 
depending on the thickness of the platelet single-crystal specimen used for the measurements. 
The superconductivity was similarly observed to disappear at lower pressure 
of $\sim$5 GPa when less hydrostatic PTM (glycerin) was used, suggesting that 
uniaxial compression along the [001] direction promotes the transition into the 
ortho I\hspace{-1pt}I phase. 

Single-crystal specimens were obtained by a chemical vapor transport method,\cite{bohmer2,bohmer3}
details of which are available in the literature.\cite{miyoshi20}
The XRD pattern of the FeSe single crystals is shown in Fig. 1(a), 
where the (00$l$) peaks are observed, indicating that the (00$l$) plane is exposed 
on the surface of the platelet crystals shown in the inset of Fig. 1(b). 
Figure 1(b) and 1(c) show the temperature ($T$) dependences of the DC magnetization ($M$) 
and electrical resistivity ($\rho$), respectively. These data indicate that 
$T_{\rm c}$ is $\sim$9 K and the nematic transition temperature $T_{\rm s}$ is $\sim$90 K 
at ambient pressure, both of which are consistent with those in earlier reports.\cite{sun,bohmer3} 
High-pressure magnetic measurements were done by using a miniature
DAC,\cite{mito} which was combined with a sample rod of 
a commercial SQUID magnetometer. We used a CuBe gasket with a 0.3 mm$\phi$ gasket hole, 
where a platelet FeSe single crystal was loaded parallel to 
the culet plane of the diamond anvil together with a small piece of high-purity Pb 
to realize the in-situ determination of pressure. The magnetization data for FeSe and Pb were 
obtained by subtracting the magnetic contribution of the DAC measured in an empty run 
from the total magnetization.\cite{miyoshi08,miyoshi09,miyoshi13,miyoshi14}
As the PTM, we mainly used Ar, which solidifies at $P$=1.2 GPa at room temperature 
as a soft molecular solid and is known to be a hydrostatic PTM.\cite{finger,tateiwa}
For the measurements, we adopted platelet single crystals with different thicknesses, which 
were $\sim$20 $\mu$m or less than $\sim$8 $\mu$m, 
while the gasket thickness was typically $\sim$30 $\mu$m (40 $\mu$m) after (before) 
the measurements.
The thickness ($t$) of the platelet specimens for $t$ $\lesssim$ 8 $\mu$m was estimated 
by using a scanning electron microscope.
Electrical resistivity of a FeSe single crystal with dimensions of 0.9$\times$0.5$\times$0.035 mm$^3$ 
was measured under pressure by using an opposed-anvil cell.\cite{kitagawa}
\begin{figure}[h]
\includegraphics[width=7.5cm]{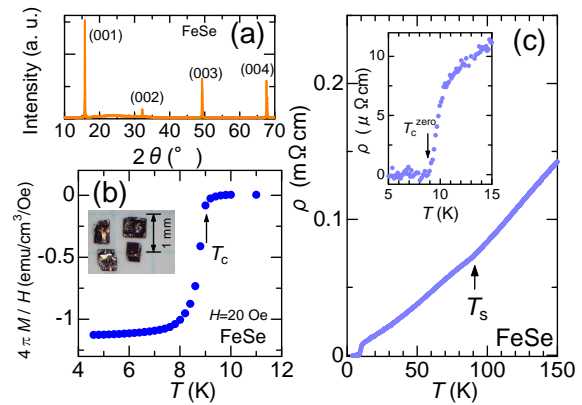}
\caption{(Color on line) X-ray diffraction pattern at room temperature (a), 
Temperature ($T$) dependence of zero-field-cooled DC magnetization $M$ measured by applying 
a magnetic field of 20 Oe parallel to the (001) plane (b) 
and electrical resistivity $\rho$ (c) 
for FeSe single crystals. The insets in (b) and (c) show 
a photographic image of typical single crystals of FeSe and 
an enlarged view of $\rho$($T$) at nearly zero resistivity, respectively. 
}
\label{autonum}
\end{figure}

Measurements were also performed using glycerin as the PTM to observe the effect 
of the degradation of the hydrostaticity. 
The $M$($T$) curves for a single crystal of FeSe with $t$$\sim$20 $\mu$m 
at various pressures measured using glycerin as the PTM are shown in Fig. 2(a). 
The curve at ambient pressure shows a sharp decrease 
below $T_{\rm c}^{\rm dia}$$\sim$9 K, and $T_{\rm c}^{\rm dia}$ increases with increasing 
pressure up to $\sim$25 K at 5.1 GPa. However, above the pressure, 
$T_{\rm c}^{\rm dia}$ decreases and the diamagnetic amplitude decreases rapidly. 
The pressure evolution of $M$($T$) 
suggests the disappearance of superconductivity above 5$-$6 GPa. 
Figure 2(b) shows the $\rho$($T$) curves measured using glycerin at various pressures. 
In Fig. 2(b), the $\rho$($T$) curve at 2.0 GPa exhibits a maximum at $T_{\rm c}^{\rm onset}$$\sim$25 K 
and a sharp decrease showing zero-resistivity below $T_{\rm c}^{\rm zero}$$\sim$15 K. 
$T_{\rm c}^{\rm zero}$ begins to decrease above 4.6 GPa and reaches 5 K at 6.4 GPa, 
while $T_{\rm c}^{\rm onset}$ is $\sim$37 K at 6.4 GPa, 
indicating a significant broadening of the superconducting transition. 
The AFM transition temperature $T_{\rm m}$ on the $\rho$($T$) curves is determined in a 
similar manner as in the literature.\cite{sun} These results are summarized in Fig. 2(c), 
where the transition temperatures are plotted as a function of pressure together with the phase 
boundary lines (broken lines) obtained from the measurements under hydrostatic pressure.\cite{sun} 
Here, both $T_{\rm c}^{\rm zero}$ and $T_{\rm c}^{\rm dia}$ 
suddenly drop above $\sim$5 GPa, thereby deviating from the broken line. 
\begin{figure}[h]
\includegraphics[width=7.5cm]{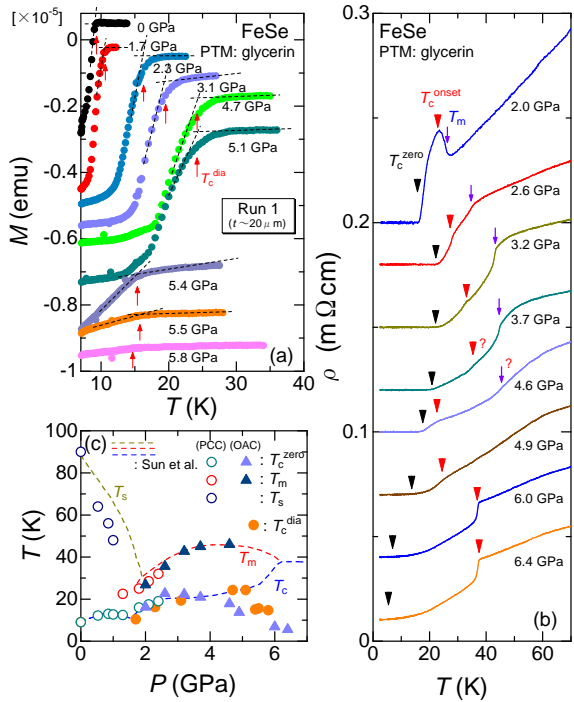}
\caption{(Color on line) Temperature dependence of zero-field-cooled DC magnetization $M$ in a 
magnetic field of 20 Oe (a) and electrical resistivity $\rho$ (b) 
at various pressures using glycerin as the PTM for FeSe. 
The data are intentionally shifted along 
the longitudinal axis for clarity. (c) Plots of diamagnetic onset ($T_{\rm c}^{\rm dia}$), 
zero-resistive ($T_{\rm c}^{\rm zero}$), nematic transition ($T_{\rm s}$) and 
AFM transition temperatures ($T_{\rm m}$). The data of $T_{\rm c}^{\rm zero}$, $T_{\rm m}$ and 
$T_{\rm s}$ obtained using a piston cylinder cell (PCC) in a previous study\cite{miyoshi20} are also plotted, 
in addition to those obtained using an opposed anvil cell (OAC). The broken lines were reproduced from the 
literature.\cite{sun}. 
}
\label{autonum}
\end{figure}

The disappearance of the superconductivity above $\sim$5 GPa 
seen in Fig. 2(a)-2(c) appears to correspond to the solidification of glycerin above 5 GPa at room 
temperature,\cite{rzoska} which induces a nearly uniaxial compression perpendicular 
to the (001) plane. It should be noted that 
the disappearance has never been observed by applying a uniaxial stress in the [101] direction 
by using NaCl as the PTM, as in the previous study.\cite{miyoshi14} 
Thus, uniaxial stress, especially along the [001] direction is likely to 
be responsible for the disappearance of the superconductivity. 
$M$($T$) curves of FeSe at various pressures for FeSe with 
$t$$\sim$20 $\mu$m and $t$$\sim$8 $\mu$m using 
Ar as the PTM, are shown in Figs. 3(a) and 3(b), respectively. Fig. 3(a) 
shows that $T_{\rm c}^{\rm dia}$ exhibits a 
local maximum at 0.82 GPa but shows a rapid increase above 1.5 GPa, reaching 
$T_{\rm c}^{\rm dia}$$\sim$30 K at 6.7 GPa. Above 6.9 GPa, the amplitude of the diamagnetic 
response decreases steeply and $M$($T$) finally almost stabilizes at 7.6 GPa. 
In contrast, the $M$($T$) curves for $t$$\sim$8 $\mu$m in Fig. 3(b) 
indicate that $T_{\rm c}^{\rm dia}$ 
exhibits a similar variation with pressure compared with that for $t$$\sim$20 $\mu$m 
up to 6.7 GPa; however, the amplitude of the diamagnetic response appears not to be reduced 
even above 7.0 GPa, yielding $T_{\rm c}^{\rm dia}$$\sim$37 K above 7.6 GPa. 
\begin{figure}[h]
\includegraphics[width=8.5cm]{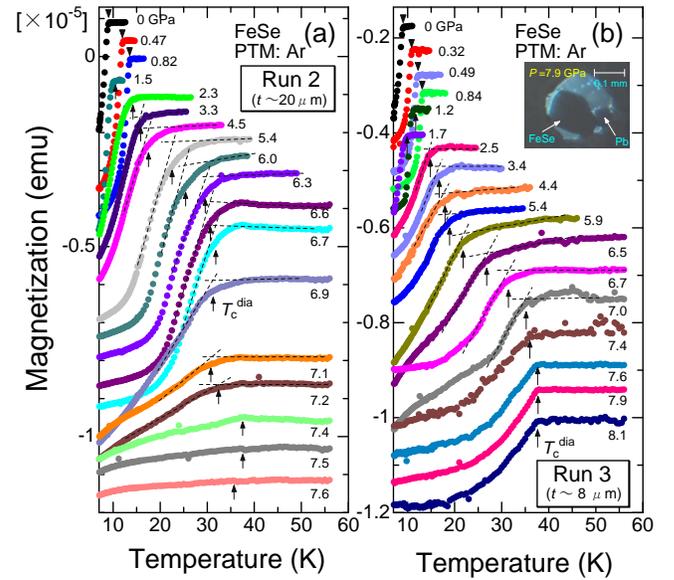}
\caption{(Color on line) (Color on line) Temperature dependence of zero-field-cooled DC magnetization for FeSe in a 
magnetic field of 20 Oe under various pressures using Ar as the PTM for a sample with thickness 
$t$$\sim$20 $\mu$m (a) and $\sim$8 $\mu$m (b). The data were intentionally shifted along 
the longitudinal axis for clarity. The inset in (b) shows the sample assembly for Run 3 at $P$=7.9 GPa. 
}
\label{autonum}
\end{figure}
\begin{figure}[h]
\includegraphics[width=8.5cm]{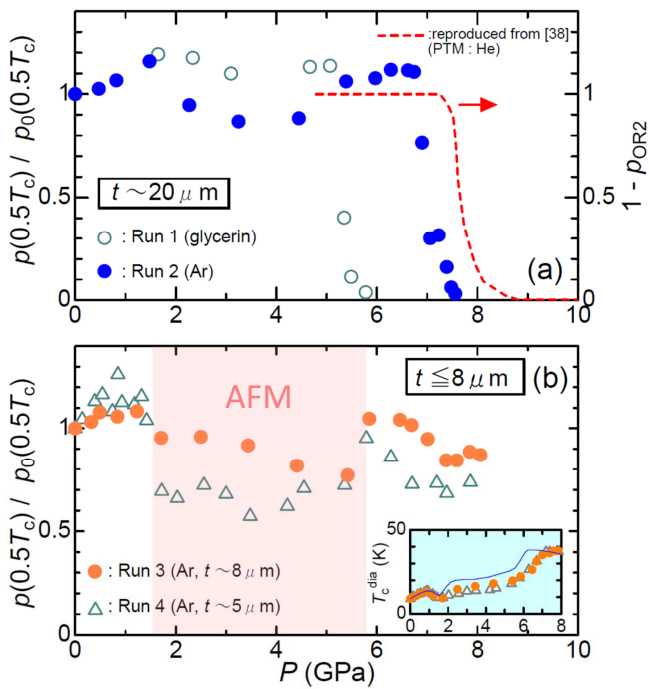}
\caption{(Color on line) (a) Superconducting volume fraction ($p$)
at $T$=0.5$T_{\rm c}$ normalized to that at ambient pressure ($p_0$) plotted versus pressure 
for FeSe single crystals with thickness $t$$\sim$20 $\mu$m. 
The broken line indicates the pressure variation of 
1$-$$p_{\rm OR2}$, where $p_{\rm OR2}$ is the phase fraction 
of the ortho I\hspace{-1pt}I phase reproduced from the literature\cite{bohmer4}.
(b) $p$(0.5$T_{\rm c}$)$/$$p_0$(0.5$T_{\rm c}$) plotted as a function of pressure for $t$ $\lesssim$ 8 $\mu$m. 
Inset: $T_{\rm c}^{\rm dia}$ versus pressure for specimens with $t$ $\lesssim$ 8 $\mu$m. 
The $T_{\rm c}$$-$$P$ curve determined mainly from the zero-resistive temperature 
by using a CAA\cite{sun} is shown as a solid line. 
}
\label{autonum}
\end{figure}
To clarify the pressure evolution of the superconducting volume fraction, we estimated it 
from the amplitude of the diamagnetic response as in the previous study.\cite{miyoshi14} 
The superconducting volume fraction ($p$) at $T$=0.5$T_{\rm c}$ normalized 
by that at ambient pressure ($p_0$), as a function of 
pressure for $t$$\sim$20 $\mu$m and $t$$\sim$8 $\mu$m, is plotted in Fig. 4(a) and 4(b), respectively. 
We excluded the contribution of Pb to the amplitude of the magnetic response 
to enable us to precisely estimate $p$(0.5$T_{\rm c}$)$/$$p_0$(0.5$T_{\rm c}$) at low pressures. 
In Fig. 4(a), $p$(0.5$T_{\rm c}$)$/$$p_0$(0.5$T_{\rm c}$) for $t$$\sim$20 $\mu$m shows an abrupt 
decrease to zero above $\sim$5 GPa when glycerin was used as the PTM, whereas a sharp decrease 
was observed at pressures above $\sim$7 GPa when Ar was used, indicating the disappearance 
of the superconductivity. The pressure variations are qualitatively similar to that of 
1$-$$p_{\rm OR2}$, where $p_{\rm OR2}$ is the phase fraction of the ortho I\hspace{-1pt}I phase and 
is evaluated by the relative intensity of Bragg reflections for the ortho I\hspace{-1pt}I 
and tetragonal phases in the XRD measurements under pressure using He as the PTM.\cite{bohmer4} 
This suggests that the origin of the disappearance of the superconductivity 
is due to the appearance of the ortho I\hspace{-1pt}I phase. 
In contrast, $p$(0.5$T_{\rm c}$)$/$$p_0$(0.5$T_{\rm c}$) for $t$$\sim$8 $\mu$m in Fig. 4(b) 
remains $\sim$0.9 even above 7 GPa. To verify the reproducibility, the measurements 
were also performed using a single crystal of another batch 
with $t$$\sim$5 $\mu$m (Run 4). The results are shown in Fig. 4(b). It is confirmed that 
$p$(0.5$T_{\rm c}$)$/$$p_0$(0.5$T_{\rm c}$) for $t$$\sim$5 $\mu$m remains more than 0.7 above 7 GPa. 
We can also observe a common tendency that 
$p$(0.5$T_{\rm c}$)$/$$p_0$(0.5$T_{\rm c}$) is slightly reduced between 1.5$\lesssim$$P$$\lesssim$6.0 GPa. 
This behavior is probably the result of the coexistence of the AFM and superconducting phases and is regarded 
as an evidence of the appearance of the AFM phase in the pressure region. 
Nonetheless, $p$(0.5$T_{\rm c}$)$/$$p_0$(0.5$T_{\rm c}$) possibly reflects the broadening of the 
superconducting transition due to the co-existing AFM phase rather than the 
reduction of the superconducting volume fraction. 
A similar weakening of the diamagnetic shielding in the pressure region where the AFM dome 
appears has been observed for FeSe$_{\rm 1-x}$S$_{\rm x}$.\cite{yip} 
In Figs. 4(a) and 4(b), a small increase is seen in $p$(0.5$T_{\rm c}$)$/$$p_0$(0.5$T_{\rm c}$) for 
0$\lesssim$$P$$\lesssim$1 GPa. This is probably attributable to an initial shrinkage 
in the thickness of the single crystal, 
which leads to an increase in the demagnetization coefficient. 
The inset in Fig. 4(b) shows the pressure variations of 
$T_{\rm c}^{\rm dia}$ for $t$ $\lesssim$ 8 $\mu$m. The $T_{\rm c}^{\rm dia}$$-$$P$ curves are 
qualitatively similar to the $T_{\rm c}$$-$$P$ curve determined 
mainly from the zero-resistive temperature by using a CAA.\cite{sun} 
 
We should note that He was used as the PTM in the previous observation of 
the ortho I\hspace{-1pt}I phase by the XRD measurements above 8 GPa 
at room temperature.\cite{svitlyk,bohmer4} 
Based on the facts, one may consider that 
the ortho I\hspace{-1pt}I phase has been induced under hydrostatic pressure, because 
He is generally known not to solidify up to $\sim$12 GPa at room temperature.\cite{bell}
However, to explain the disappearance and survival of the superconductivity, 
we propose a possible scenario, in which the ortho I\hspace{-1pt}I phase appears 
owing to the degradation of the hydrostaticity of the pressure for $t$$\sim$20 $\mu$m, 
while the ortho I\hspace{-1pt}I phase does not appear for $t$ $\lesssim$ 8 $\mu$m 
because of the improvement in the hydrostaticity resulting from 
the decrease in the thickness of the crystal. The results for $t$ $\lesssim$ 8 $\mu$m are 
similar to those obtained in the measurements using a CAA, which 
generates the perfect hydrostatic pressure by three-axis compression. 
Indeed, it has been shown in XRD measurements of Au using a DAC with He
that the increase in the uniaxial stress component above 30 GPa 
tends to be suppressed as the height of the sample decreases 
along the load direction.\cite{takemura} 
In addition, we failed to detect any sign 
of the disappearance of the superconductivity using Ar as the PTM in the previous 
observation for a single crystal with dimensions of $\sim$0.1$\times$0.1$\times$0.02 mm$^3$
where the (101) plane is exposed on the surface.\cite{miyoshi14} 
This suggests that the disappearance is induced by a uniaxial component of the compression 
along a specific direction [001]. This conclusion is also supported by the results 
shown in Figs. 2(a)-2(c), where solid-state compression along the [001] direction 
accelerates the transformation to the ortho I\hspace{-1pt}I phase. 
In fact, it was shown in a previous study that a large shrinkage along the c-axis 
in the crystal structure is accompanied by the transition to the ortho I\hspace{-1pt}I 
phase, yielding a volume reduction of 11$\%$ at 300 K.\cite{svitlyk} This suggests that 
a uniaxial stress along [001] direction could be important for the occurrence of the transition. 
Thus, the above-mentioned scenario is most likely, 
although it is unclear how 
uniaxial stress can be generated for $P$ $<$ 12 GPa at room temperature by using He as PTM. 
The relationship among the three phases on the $T$-$P$ phase diagram still 
remains to be fully understood not only in pure-FeSe, 
but also in S-doped\cite{matsuura} and Te-doped FeSe,\cite{mukasa} 
as is suggested by the recent study reporting the coexistence of 
superconductivity and magnetic order on a tetragonal lattice above 6 GPa.\cite{bohmer4} 
Further investigations at high pressure are necessary in future studies. 
However, special attention should be given to the hydrostaticity of the pressure, 
which greatly affects the crystal structure of 
the single crystal specimen with the (001) surface. 

In summary, by conducting the DC magnetization measurements, 
we successfully observed the disappearance of the superconductivity in FeSe owing to the 
appearance of the ortho I\hspace{-1pt}I phase. In addition, the survival of the superconductivity 
owing to the absence of the ortho I\hspace{-1pt}I phase was also observed using 
platelet single crystal specimens with different thicknesses. 
It is inferred from the results using both Ar and glycerin for the specimens 
with $t$$\sim$20 $\mu$m that uniaxial stress along the [001] direction is essential 
to realize the ortho I\hspace{-1pt}I phase. 
The superconducting volume fraction for specimens with $t$ $\lesssim$ 8 $\mu$m was found to show 
no sudden decrease at high pressure. This is consistent with the result observed 
in the measurements under hydrostatic 
pressure using a CAA\cite{sun}, such that the hydrostaticity 
can be improved by reducing the thickness of the specimens in our measurements. 
In conclusion, we emphasize that the thickness of the sample can be important for the hydrostaticity 
in the measurements using a DAC, even though a hydrostatic PTM is used. 

\begin{acknowledgments}
This work was supported by JSPS KAKENHI (Grant Number JP18K03516). 
The authors thank T. Matsumoto, D. Morii, T. Ohyama, A. Nishiyama and T. Sueyasu 
for technical assistance. 

\end{acknowledgments}

\end{document}